\documentclass[journal,10pt]{IEEEtran}

\usepackage{CJK}
\usepackage{cite}
\usepackage[justification=centering]{caption}
\usepackage{graphicx}
\usepackage{epstopdf}
\usepackage{subfigure}
\usepackage[cmex10]{amsmath}
\usepackage[amsmath,thmmarks]{ntheorem}
\usepackage{amssymb}
\usepackage{ntheorem}
\usepackage{mathabx}
\usepackage{algorithm,float}
\usepackage{algorithmic}
\usepackage{lipsum}
\usepackage{booktabs}
\usepackage{threeparttable}
\usepackage{multicol}
\usepackage{multirow}
\usepackage{xcolor}
\usepackage{bm}
\usepackage{mathabx}
\usepackage{stmaryrd}
\usepackage{upgreek}
\usepackage{url}
\usepackage{wasysym}
\usepackage[labelsep=none]{caption}
\usepackage{lettrine}

\newtheorem{myDef}{Definition}

\begin{document}

\title{Multi-Agent Reinforcement Learning for Joint Cooperative Spectrum Sensing and Channel Access in Cognitive UAV~Networks}

\author{Weiheng Jiang,~\IEEEmembership{Member,~IEEE}, Wanxin Yu, Wenbo Wang,~\IEEEmembership{Member,~IEEE}, Tiancong Huang

\thanks{Weiheng Jiang, Wanxin Yu and Tiancong Huang are with the School of Microelectronics and Communication Engineering, Chongqing University, Chongqing,~China,
(email: whjiang@cqu.edu.cn, wxyuwan@cqu.edu.cn, htc@cqu.edu.cn).}
\thanks{Wenbo Wang is with the Faculty of Engineering, Bar Ilan University, Ramat Gan, Israel,
(email: wangwen@biu.ac.il).}}

{}

\maketitle

\begin{abstract}
This paper studies the problem of distributed spectrum/channel access for cognitive radio-enabled unmanned aerial vehicles (CUAVs) that overlay upon primary channels. Under the framework of cooperative spectrum sensing and opportunistic transmission, a one-shot optimization problem for channel allocation, aiming to maximize the expected cumulative weighted reward of multiple CUAVs, is formulated. To handle the uncertainty due to the lack of prior knowledge about the primary user activities as well as the lack of the channel-access coordinator, the original problem is cast into a competition and cooperation hybrid multi-agent reinforcement learning (CCH-MARL) problem in the framework of Markov game (MG). Then, a value-iteration-based RL algorithm, which features upper confidence bound-Hoeffding (UCB-H) strategy searching, is proposed by treating each CUAV as an independent learner (IL). To address the curse of dimensionality, the UCB-H strategy is further extended with a double deep Q-network (DDQN). Numerical simulations show that the proposed algorithms are able to efficiently converge to stable strategies, and significantly improve the network performance when compared with the benchmark algorithms such as the vanilla Q-learning and DDQN algorithms.
\end{abstract}


\IEEEpeerreviewmaketitle

\section{Introduction}

Recent years have witnessed remarkable success of unmanned aerial vehicle (UAV) clusters in a variety of scenarios ranging from disaster relief to commercial applications of unmanned swarm operations \cite{ye2020dynamic,wu2020cellular}. As one backbone technology for UAV systems, communication protocol design for UAVs, thus, naturally receives intensive attention from both academia and industry \cite{ma2020impact,jingnan2017research}. However, due to the ad hoc nature of UAV networks, directly applying the off-the-shelf wireless access protocols for vehicle-to-vehicle (V2V) becomes a difficult task, especially when the UAVs have to overlay upon the spectrum occupied by an existing infrastructure and ensure zero interference. In this regard, the adoption of cognitive radio (CR) technologies \cite{liu2020reinforcement,ning2020reinforcement} into UAV systems becomes a tempting solution, since it not only avoids a series of problems caused by the rigid fixed-spectrum authorization model \cite{liu2020reinforcement,ning2020reinforcement}, but also has the potential to adapt to a complex and time-varying radioactive environment. Nevertheless, UAVs are typically constrained by their on-device computation capabilities, but are required to quickly respond to the radio environment changes with limited coordination. Therefore, designing an intelligent mechanism to efficiently perform spectrum sensing and distributed channel access becomes a challenge of vital~importance.

So far, pioneering studies have established a number of different frameworks for spectrum sensing in CR networks \cite{xu2018efficient,shen2019uav,nie2019max}. For instance, in \cite{xu2018efficient}, an iterative signal compression filtering scheme is proposed to improve the spectrum sensing performance of CR-enabled UAV (CUAV) networks. Its core idea is to adaptively eliminate the primary user (PU) component in the identified sub-channel, and directly update the measured value to detect other active users. In \cite{shen2019uav}, the space--time spectrum sensing problem for CUAV network in the three-dimensional heterogeneous spectrum space is discussed. Spectrum detection is improved based on the fusion of sensing results over both the time domain and the space domain. In \cite{nie2019max}, aiming to reflect the dynamic topology change of the CUAV network, a clustering method based on the maximum and minimum distances of nodes is proposed to improve the performance of cooperative spectrum sensing. With these spectrum sensing methods, channel access schemes, such as channel rendezvous for opportunistic channel reservation \cite{feng2019cogmor} and one-shot optimization-based channel allocation \cite{liang2020throughput}, can be deployed for throughput-optimal allocation for CUAVs.

The above studies tackle the channel sensing and allocation problem in CUAV networks by assuming that the radioactive environment is static, and a centralized information aggregator (e.g., a leader UAV) exists. {However, more than frequently, a practical CUAV network not only faces a time-varying channel environment, i.e., the UAV--ground communication with multiple antennas will incur a 3D nonstationary geometry-based stochastic channel \cite{iet2019} or the ultra-wideband communication with the Saleh-Valenzuela time-varying statistical channel model \cite{9119755}, but is also deployed in an ad hoc manner. Therefore, it is necessary to develop a distributed sensing--allocation mechanism that causes an affordable level of overhead due to V2V information exchange.} {For this reason, a series of distributed allocation mechanisms, in particular, based on reinforcement learning (RL), are proposed to replace the traditional self-organizing schemes for spectrum sensing or channel access \cite{ning2020reinforcement,lunden2013multiagent,chen2016joint,mnih2015human,zhang2019multi,li2020deep}}. In \cite{ning2020reinforcement}, a novel Q-learning-based method is proposed for secondary users (SUs) to select cooperative sensing nodes using the discounted upper confidence bound (D-UCB) for strategy exploration and reducing the number of sensing samples. In~\cite{lunden2013multiagent}, a neighbor-based cooperative sensing mechanism using Q-learning is proposed for collaborative channel sensing by SUs. In \cite{chen2016joint}, a robust joint sensing--allocation scheme is proposed based on RL to counter the impact of adversary SUs (e.g., spectrum sensing data falsification attackers). Compared with these tabular-search-based RL methods, deep neural networks (e.g., deep Q-network) are adopted for state-value approximation \cite{mnih2015human}. Still, for the cooperative spectrum sensing problem \cite{zhang2019multi}, a multi-agent deep reinforcement learning method was adopted, and each secondary user learns an efficient sensing strategy from the sensing results to avoid interference to the primary users, also in which the upper confidence bound with Hoeffding-style bonus is used to improve the efficiency of exploration. Furthermore, cooperative multi-agent RL (MARL) methods are proposed for dynamic spectrum sensing and aggregation \cite{li2020deep}, typically with the aim of maximizing the number of successful transmissions without interrupting PUs.

In summary, most of the existing studies treat the problems of high-precision spectrum sensing and dynamic access channel allocation separately. {However, how to jointly optimize the cooperative channel sensing and spectrum access processes remains open issues, especially in the time-varying radio environment.  In addition, although a plethora of distributed algorithms (some based on RL \cite{cai2020coordination,lo2013reinforcement,zhang2018distributed}) have been proposed in the literature,} most of them are subject to rigid assumptions and cannot be directly adopted by CUAV applications, which, for example, usually emphasize network/spectrum scalability or face real-world constraints such as limited sensing/signaling capabilities and limited energy/computation resources. These concerns naturally lead to the consideration of formulating the joint sensing-and-access problem from the perspective of UAVs. As a result, the decision process of CUAVs may face more complex coupling problems in terms of sensing-and-access strategies when compared with the purely cooperative methods. Based on these considerations, this paper investigates the semicompetitive channel-sensing-and-access problem in CUAV networks where the spectrum sensing phase is organized cooperatively based on the exchange of binary sensing results. An MARL-based framework of strategy searching is proposed in the form of two distributed execution algorithms that address state-value representation differently. {The main contribution of this paper is summarized as~follows:}

\begin{itemize}
\item[$\bullet$] To coordinate the behaviors of various CUAVs for efficient utilization of idle spectrum resources of PUs, a CUAV channel exploration and utilization protocol framework based on sensing--fusion--transmission is proposed.
\item[$\bullet$] A problem maximizing the expected cumulative weighted rewards of CUAVs is formulated. Considering the practical constraints, i.e., the lack of prior knowledge about the dynamics of PU activities and the lack of a centralized access coordinator, the original one-shot optimization problem is reformulated into a Markov game (MG). A weighted composite reward function combining both the cost and utility for spectrum sensing and channel access is designed to transform the considered problem into a competition and cooperation hybrid multi-agent reinforcement learning (CCH-MARL)~problem.
\item[$\bullet$]To tackle the CCH-MARL problem through a decentralized approach, UCB-Hoeffding (UCB-H) strategy searching and the independent learner (IL) based Q-learning scheme are introduced. More specifically, UCB-H is introduced to achieve a trade-off between exploration and exploitation during the process of \emph{Q}-value updating. Two decentralized algorithms with limited information exchange among the CUAVs, namely, the IL-based Q-learning with UCB-H (IL-Q-UCB-H) and the double deep Q-learning with UCB-H (IL-DDQN-UCB-H), are proposed. The numerical simulation results indicate that the proposed algorithms are able to improve network performance in terms of both the sensing accuracy and channel utilization.
\end{itemize}

The rest of this paper is organized as follows. Section \ref{sec2} presents the network model and formulates the problem from a centralized perspective. Section \ref{sec3} casts the problem into the context of MARL, and Section \ref{sec4} proposes the RL-based solutions for joint spectrum sensing and channel access. Simulation results and analyses are presented in Section \ref{sec5}. Section \ref{sec6} concludes the paper.


\section{System Model}\label{sec2}
\subsection{Network Model}

\begin{figure}
\includegraphics[width=1.0\linewidth]{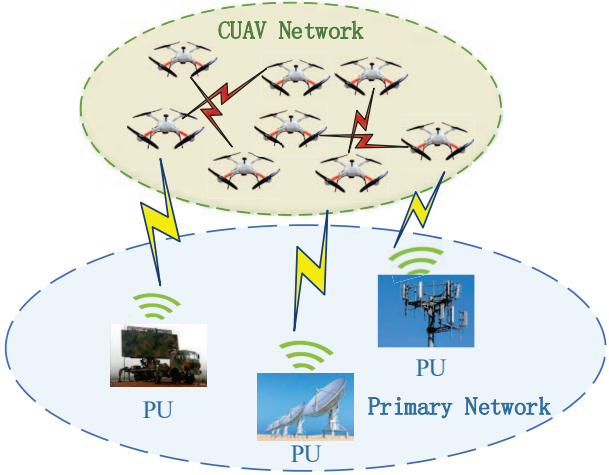}
\caption{$~$The network structure of CUAVs coexisting with PUs.\label{Figure1.}}
\end{figure}

Consider a coexistence network scenario, as shown in Fig.~\ref{Figure1.}, where a cluster of $N$ CUAVs try to access $M$ orthogonal primary spectrum resources in an overlaying mode over the airspace of interest.  Herein, the cluster CUAVs perform cooperative area sensing and data backhaul tasks (e.g., geological survey, target monitoring, etc.) \cite{kaur2020energy,nobar2021resource}. The CUAVs perform cooperative spectrum sensing to opportunistically exploit the idle spectrum resources of the primary users (PUs). For our considered CUAV network, since the communication demands are mainly from the task cooperation among the CUAVs, the communication channels used by CUAVs are dominated by the line-of-sight (LoS) air-to-air (A2A) channels \cite{cui2019multi}. Meanwhile, due to the platooning characteristics of the CUAV cluster, the communication channels between any two CUAVs can be treated as quasi-static over the task period \cite{chandrasekharan2016designing}.

\begin{figure}
\includegraphics[width=1.0\linewidth]{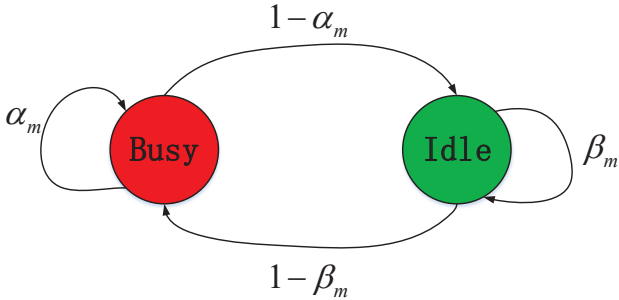}
\caption{$~$Occupancy state transition diagram of PU channel $m$.\label{Figure2.}}
\end{figure}

Due to the limit on hardware capabilities, we consider that a CUAV performs narrow-band spectrum sensing and can sense and access at most one single PU channel at a given time slot \cite{lunden2013multiagent,zhang2019multi}. Meanwhile, it is possible that not all of the active PUs are within the sensing range of all the CUAVs. This results in poor reliability of the sensing result by a single CUAV, and thus cooperative sensing is desired for CUAVs to improve the sensing performance collectively. Furthermore, we assume that the PU networks over different target frequency bands provide heterogeneous services to their users, such as data communications, radar, or other dynamic spectrum occupancy services. The heterogeneous channel bandwidth of PU channel $m$ is denoted by $B_{m}$. In addition, we assume that PU services are bursty, and can be described by a slotted (discrete-time) Markov process of two states (i.e., busy and idle) \cite{li2020deep} as shown in Fig.~\ref{Figure2.} with a pair of state transition probabilities $(\alpha_m,\beta_m)$.

\subsection{Framework of Channel Sensing and Access}
\begin{figure}
\includegraphics[width=1.0\linewidth]{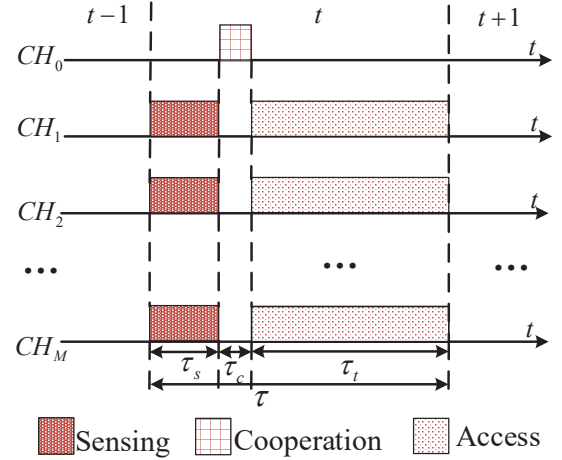}
\caption{$~$Structure of one time slot for the joint channel sensing and access protocol.\label{Figure3.}}
\end{figure}

To enable the coexistence of multiple CUAVs over a limited number of PU frequency bands, we need a protocol framework to coordinate the channel sensing-and-access behaviors of these CUAVs. We assume that the CUAVs are able to access and synchronize over a dedicated common control channel (CCC), i.e., $CH_0$ in Fig.~\ref{Figure3.}, which the spectrum sensing results and channel selection decisions can be shared among the CUAVs. We also assume that the CUAVs operating on the same PU channel transmit with nonorthogonal spectrum sharing techniques. The processes of spectrum sensing and channel access are organized in time slots (see Fig.~\ref{Figure3.}). More specifically, each time slot of PU channel sensing and utilization by CUAVs is divided into three consecutive sub-frames of sensing ($\tau_s$), cooperation ($\tau_c$), and access/transmission ($\tau_t$). At the beginning of the sub-frame of sensing, CUAVs decide on which channels to sense and access by switching their transceiver operations to the corresponding channels. Note that in this sub-frame, some of the CUAVs may stay idle and select no channel.  In the subsequent sub-frame of cooperation, each CUAV broadcasts their own sensing results over the CCC in an orderly manner. Based on the received sensing results, each CUAV is able to perform the local sensing-result fusion and obtain a uniform vector of state observation as the other CUAVs. The local fusion results will be used for deciding on whether to access or not in the last sub-frame of access.

We assume that the messages exchanged over the CCC are reliable (cf. \cite{zhang2019multi}), and for cooperation, we assume that the same fusion rule, such as the ``K-out-of-N'' or ``AND'' rules~\cite{chen2011cooperative,han2010efficient}, is adopted by all the CUAVs. This ensures that all the CUAVs obtain a consistent observation about the status (i.e., busy or idle) of PU channels. Obviously, the more CUAVs participating in sensing the same channel, the higher accuracy of the sensing result is \cite{abdi2015optimum}. However, since the CUAVs choose to access the same channel that they sense, this will also lead to a higher congestion level over the PU channel. Therefore, the CUAVs need to develop a proper channel selection strategy to balance between the spectrum sensing accuracy (i.e., to reduce transmission failure probability) and the quality of transmissions (i.e., to avoid severe congestion over the selected channel).

\subsection{Problem Formulation}
Given the presented network model and the proposed access protocol, we know that the network performance is determined by the channel selection strategies of the CUAVs for joint channel sensing-and-access. Our goal is to find an appropriate approach to jointly reflect the system cost in cooperative spectrum sensing and the utility in successful transmissions. Furthermore, we aim to derive an optimal joint strategy of the CUAVs for channel selection in the time-varying radio environment, such that the utility of PU channels is maximized. Therefore, from a Genie's perspective, we can formulate the following centralized optimization problem for the considered CUAV network:
\begin{equation}
\begin{aligned}
\max\limits_{\{c_{n,m}^t\}}\,&\mathbb{E}\left[\sum_{t=0}^T\sum_{n=1}^N\sum_{m=0}^M c_{n,m}^t\gamma^tr_{n,m}^t\right] \\
\textrm{s.t.} &\sum_{m=0}^M c_{n,m}^t\leq 1,\forall n=1,\ldots,N, \\
&c_{n,m}^t\in\{0,1\},
\label{eq:1}
\end{aligned}
\end{equation}
where $T$ is the total number of time slots for CUAV network operation. In practice, $T$ is typically not known in advance. $c_{n,m}^t$ is the binary decision variable of CUAV $n$ on PU channel $m$ in time slot $t$, and $c_{n,m}^t=1$ if CUAV $n$ selects PU channel $m$ to sense and access at time slot $t$. $r_{n,m}^t$ is the reward of CUAV $n$ on PU channel $m$ at time slot $t$, and is determined by the weighted sum of user sensing access cost and utility. For ease of discussion, we defer the detailed definition of $r_{n,m}^t$ to Section \ref{sec3}. Obviously, we have $r_{n,m}^t=0$ if $c_{n,m}^t=0$. $\gamma\in\left[0,1\right)$ is the reward discount factor to translate the future rewards into the reward at $t=0$ \cite{sutton2018reinforcement}.

In addition, the expectation operation $\mathbb{E}\left[\cdot\right]$ is calculated over the PU channel evolution model (see also Fig.~\ref{Figure2.}). Without considering the expectation operation $\mathbb{E}[\cdot]$, (\ref{eq:1}) will degrade to be an one-shot, NP-hard binary programming problem. However, in the real world, the PU channel evolution model is not known in advance, and it is impractical to assign a centralized coordinator in the CUAV cluster, due to the constraints of on-device computation/signaling capability. Therefore, in the following, we reformulate the static problem as described in (\ref{eq:1}) into a CCH-MARL problem based on MG, and then resort to the MARL-based algorithms for deriving the channel selection strategies of the CUAVs.

\section{Problem Modeling Based on MARL}\label{sec3}
\subsection{Markov Game-Based Problem Formulation}\label{sec31}

Before proceeding to the reformulation of the considered problem, we provide the definition of MG as follows.

\begin{myDef}[Markov game \cite{zhang2021multi}]
\label{Definition_1}
An MG is defined by a sixtuplet as\\ $\langle\mathcal{N},\mathcal{S},\{\mathcal{A}_n\}_{n\in{\mathcal{N}}},
\mathcal{P},\{{r}_n\}_{n\in{\mathcal{N}}},\gamma\rangle$, where
\begin{itemize}
    \item $\mathcal{N}=\{1,\ldots,N\}$ is the set of agents.
    \item $\mathcal{S}$ is the state space observed consistently by all agents.
    \item $\mathcal{A}_n$ is the action space of agent $n$, and the joint action space of all the agents is $\mathcal{A}:=\mathcal{A}_1\times\cdots\mathcal{A}_N$.
    \item $\mathcal{P}:\mathcal{S}\times\mathcal{A}\rightarrow\Delta(\mathcal{S})$ is the transition probability from any state $\bm{s}\in\mathcal{S}$ to any state $\bm{s'}\in\mathcal{S}$ for any given joint action $\bm{a}=(a_1,a_2,...,a_n)\in\mathcal{A}$.
    \item The reward function ${r}_n:\mathcal{S}\times\mathcal{A}\times\mathcal{S}\rightarrow\mathbb{R}$ determines the instant reward received by agent $n$ in the controlled Markov process from $(\bm{s},\bm{a})$ to $\bm{s'}$.
    \item $\gamma\in[0,1]$ is the reward discount factor.
\end{itemize}
\end{myDef}

Based on \textbf{Definition~\ref{Definition_1}}, we are able to map the considered optimization problem from~(\ref{eq:1}) into the following MG:
\begin{itemize}
    \item \emph{Agent Set}
  $\mathcal{N}$ consists of the $N$ CUAVs (agents), i.e.,  $\mathcal{N}=\{1,...,N\}$. 
    \item \emph{State space} $\mathcal{S}$ of the MG is defined as
    \begin{equation}
        \mathcal{S}=\{\bm{s}^t=(s_0^t\ldots,s_M^t,o_1^t,\ldots,o_M^t)\},
        \label{eqn:state}
    \end{equation}
    where $s_m^t\in\{0,1,\ldots,N\}$ is the number of CUVAs that select PU channel $m$ to sense and access in the previous time slot. In particular, $s_0^t$ is the number of CUVAs that do not select any PU channel. Since each CUAV can select at most one single PU channel for sensing-and-access, $\sum_{m=0}^M s_m^t=N$. $o_m^t\in\{0,1\}$ is the observed occupancy state of PU channel $m$ in the previous time slot. Following (\ref{eqn:state}), the size of the state space is $|\mathcal{S}|=2^M\cdot(M+1)^N$. 
    \item \emph{Action space} $\mathcal{A}_n$ for CUAV $n$ is defined as $\mathcal{A}_n=\{0,1,\ldots,M\}$. Let $a_n^t\in\mathcal{A}_n$ denotes the PU channel selected by agent $n$ at time slot $t$, $a_n^t=0$ indicates that no channel is selected. The joint action space $\mathcal{A}=\prod_{n=1}^N\mathcal{A}_n$ can be defined as the Cartesian product of all the CUAVs, and the joint action at time slot $t$ is $\bm{a}^t=(a_1^t,\ldots,a_N^t)\in\mathcal{A}$.
    \item \emph{State transition probability} $\mathcal{P}$ consists of the transition maps $P(\bm{s}'|\bm{s},\bm{a})$ for all $\bm{s}'$, $\bm{s}$ and $\bm{a}$. Note that for the elements of transition $o_m\rightarrow o'_m$, the transition probability is determined by the two-state Markov process shown in Fig.~\ref{Figure2.}.
    \item \emph{Reward function} $r_n^{t+1}$ of CUAV $n$, is observed at time slot $t+1$ after the CUAVs taking a joint action $\bm{a}^t$. The details of the reward $r_n^{t+1}$ are presented in the next subsection.
\end{itemize}

\subsection{Definition of CUAVs' Reward Function}\label{sec32}
Let $m$ and $\mathcal{N}_m^{t+1}$ denote the PU channel selected by CUAV $n$ (i.e., $a_n^t=m$) and the CUAV set selecting the same channel at time slot $t+1$, respectively. For the considered CUAV network, the reward of each CUAV is defined by the weighted sum of the cost due to its spectrum exploration (spectrum sensing) and the utility obtained from channel utilization (channel access). The reward $r_n^{t+1}$ for CUAV $n$ is defined as

\begin{equation}
\begin{aligned}
&r_n^{t+1}(\bm{s}^{t+1},\bm{s}^{t},\bm{a}^t)=\\
&\begin{cases}
-E_{ss,n}^{t+1},&\text{if } a_n^t=m, o_{m}^{t+1}=d_{m}^{t+1}=1,\\
-E_{ss,n}^{t+1}-E_{dt,n}^{t+1},&\text{if } a_n^t=m, o_{m}^{t+1}=1,d_{m}^{t+1}=0,\\
-\eta E_{ss,n}^{t+1}-\mu E_{dt,n}^{t+1}\\
\quad\quad\quad\quad+(1-\eta-\mu)R_n^{t+1},&\text{if } a_n^t=m, o_{m}^{t+1}=d_{m}^{t+1}=0,\\
-\eta E_{ss,n}^{t+1}-(1-\eta)R_n^{t+1},&\text{if } a_n^t=m, o_{m}^{t+1}=0,d_{m}^{t+1}=1,\\
0,&\text{if }a_n^t=0,
\end{cases}
\end{aligned}
\label{eq:2}
\end{equation}
where $d_{m}^{t+1}\in\{0,1\}$ is the sensing fusion result of the cooperative CUAVs over PU channel $m$ at time slot $t+1$. $d_{m}^{t+1}$ is a function of $\bm{a}^t$, i.e., $d_{m}^{t+1}=f(\bm{a}^t)$, and the form of $f(\cdot)$ is determined by the adopted sensing fusion rule. We note that due to the inevitable missed detection and false alarm \cite{lunden2013multiagent}, the real PU channel state $o_{m}^{t+1}$ may not be consistent with the sensing fusion result $d_{m}^{t+1}$ and thus we have the first four cases in (\ref{eq:2}). In (\ref{eq:2}), $E_{ss,n}^{t+1}$ and $E_{dt,n}^{t+1}$ are the spectrum sensing and channel access cost for CUAV $n$, respectively. More specifically, the cost of sensing/access is mainly incurred by the energy consumption of the transceiver for spectrum sensing and data transmission. $R_n^{t+1}$ is the reward corresponding to the amount of successively transmitted data during time slot $t+1$ for CUAV $n$. $\eta\in(0,1)$ and $\mu\in(0,1)$ are the weighting factors for the spectrum sensing and channel access cost, respectively. The five cases in (\ref{eq:2}) are further explained as follows:
\begin{itemize}
    \item [(i)] If PU channel $m$ is busy, and the sensing fusion result is the same, i.e., $o_{m}^{t+1}=d_{m}^{t+1}=1$, the reward of CUAV $n$ is solely determined by the spectrum sensing cost $-E_{ss,n}^{t+1}$.
    \item [(ii)] If PU channel $m$ is busy but the sensing fusion result leads to a missed detection, i.e., $o_{m}^{t+1}=1,d_{m}^{t+1}=0$, CUAV $n$'s reward is determined by the sum of spectrum sensing cost $-E_{ss,n}^{t+1}$ and the cost due to the failed data transmission, $-E_{dt,n}^{t+1}$.
    \item [(iii)] If PU channel $m$ is idle and the sensing fusion result is the same, i.e., $o_{m}^{t+1}=d_{m}^{t+1}=0$, CUAV $n$'s reward is determined by the weighted sum of the sensing cost, $-E_{ss,n}^{t+1}$, the cost for data transmission, $-E_{dt,n}^{t+1}$, and the utility of successful transmission, $R_n^{t+1}$.
    \item [(iv)] If PU channel $m$ is idle but the fusion result leads to a false alarm, i.e., $o_{m}^{t+1}=0, d_{m}^{t+1}=1$, the reward of CUAV $n$ is determined by the weighted sum of spectrum sensing cost $-E_{ss,n}^{t+1}$ and the lost transmission utility $-R_n^{t+1}$.
    \item [(v)] If CUAV $n$ does not select any PU channel, i.e., $a_n^t=0$, the reward is 0.
\end{itemize}

Furthermore, we adopt the following forms of $E_{ss,n}^{t+1}$, $E_{dt,n}^{t+1}$, and $R_n^{t+1}$ in (\ref{eq:2}):
\begin{itemize}
    \item \emph{Spectrum sensing cost} $E_{ss,n}^{t+1}$ for CUAV $n$ at time slot $t+1$ is defined as the energy consumed for spectrum sensing, namely, a function proportional to the working voltage $V_{DD}$ of the receiver, the bandwidth of the sensed channel $B$, and the sensing duration $\tau_{t,n}$ \cite{zhang2012mili}:
    \begin{equation}
    E_{ss,n}^{t+1}=\tau_{t,n}V_{DD}^2B_{m}.
    \label{eq:3}
    \end{equation}
    \item \emph{Data transmission cost} $E_{dt,n}^{t+1}$ for CUAV $n$ in time slot $t+1$ is defined as the energy consumed for data transmission during the time slot,
    \begin{equation}
    E_{dt,n}^{t+1}=\tau_{s,n}p_{s,n},
    \label{eq:4}
    \end{equation}
    where $\tau_{s,n}$ and $p_{s,n}$ are the data transmission duration and transmit power, respectively. $\tau_{t,n}$, $\tau_{s,n}$, and $p_{s,n}$ are assumed to be the same for all the CUAVs, i.e., $\tau_{t,n}=\tau_t,\tau_{s,n}=\tau_s,p_{t,n}=p_t,\forall n\in\mathcal N$.
    \item \emph{Transmission utility} $R_n^{t+1}$ for CUAV $n$ in time slot $t+1$ of (cf. \emph{Cases iii} and \emph{iv}) is measured as the amount of data transmitted over the time slot. We consider that the quality of transmission is evaluated based on the throughput over a given channel under the co-channel interference:
    \end{itemize}
    \begin{equation}
    R_n^{t+1}=\tau_tB_{m}\text{log}_2(1+SINR_{n,m}^{t+1}),
    \label{eq:5}
    \end{equation}
    where $SINR_{n,m}^{t+1}$ is the received signal-to-interference-to-noise ratio (SINR) for CUAV $n$ over its selected PU channel $m$. $SINR_{n,m}^{t+1}$ can be expressed as
    \begin{equation}
    SINR_{n,m}^t=\frac{g_{n,m}p_t}{\sum_{j\in\mathcal{N}_m^t,j\ne n}g_{j,m}^np_t+\sigma^2},
    \label{eq:6}
    \end{equation}
    where $\sigma^2$ is noise power. $g_{n,m}$ is the channel gain of CUAV $n$ on PU channel $m$ and $g_{j,m}^n$ is the channel gain between CUAV $j$ and CUAV $n$ on PU channel $m$. As mentioned earlier, with platooning of the CUAV cluster, the channel gains among the CUAVs could be considered as quasi-static over the period of interest. $\sum_{j\in\mathcal{N}_m^t,j\ne n}g_{j,m}^np_t$ is the co-channel interference from the other CUAVs sharing the same PU channel $m$. Since the spatial positions and the transmitting--receiving relationship of the CUAVs over the same channel are not necessarily the same, the channel gains between different CUAVs are different, and thus the SINR of the received signals of each CUAV are different.

Finally, we examine the impact of fusion rules on the sensing fusion result $d_{m}^{t+1}=f(\bm{a}^t)$ in (\ref{eq:2}). In this paper, the ``K-out-of-N'' spectrum sensing fusion rule \cite{chen2011cooperative} is adopted to obtain the final spectrum sensing fusion result, namely,
\begin{equation}
d_m^{t+1}=\begin{cases}
1,\quad \text{if}\sum\limits_{i\in\mathcal{N}_m^{t+1}} 1_{\{d_{i,m}^{t+1}=1\}}\geqslant K,\\
0,\quad\text{others},
\end{cases}
\label{eq:7}
\end{equation}
where ${1}_{\{A=B\}}$ is the indicator function taking the value of $1$ if the condition $A=B$ is true and $0$ otherwise. Especially, it is known that for (\ref{eq:7}), if $K=1$, the ``K-out-of-N'' rule degrades to the ``OR'' rule, while if $K=N$, the ``K-out-of-N'' rule becomes the ``AND'' rule~\cite{chen2011cooperative}. We assume that the observation of each CUAV follows an independent, stationary observation process on the binary Markov process in Fig.~\ref{Figure2.}.

\subsection{MARL Algorithm Framework}

When the model of the state transition in the established MG is unknown to the CUAVs, we aim to learn to optimize the long-term statistical performance of the CUAV network. From the perspective of a single CUAV $n$, the problem of social optimization in~(\ref{eq:1}) is transformed into the following local optimization problem $\forall n\in\mathcal N$:
\begin{equation}
\max_{\pi_n}\left(v_n(\bm{s}^0,\pi_n, \bm{\pi}_{-n})=\sum_{t=0}^{+\infty}
\gamma^t\mathbb{E}(r_n^{t+1}|{\pi_n,\bm{\pi}_{-n},\bm{s}^0})\right),
\label{eq:8}
\end{equation}
where the value of the discount factor $\gamma$ reflects the effect of future rewards on optimal decision-making particularly. $\bm{\pi}_{-n}$ denotes the joint policy taken by the other CUAVs except CUAV $n$. $v_n(\bm{s}^0,\pi_n,\bm{\pi}_{-n})$ is the value function for the given state $\bm{s}^0$ and joint policy $(\pi_n,\bm{\pi}_{-n})$. Herein, the policy of CUAV $n$ is defined as $\pi_n:\mathcal{S}_n\rightarrow\Delta(\mathcal{A}_n)$, where $\Delta(\mathcal{A}_n)$ is the collection of probability distributions over CUAV $n$'s action space $\mathcal{A}_n$. $\pi_n(a_n^t|\bm{s}_n^t)$ in $\pi_n(\bm{s}_n^t)=\{\pi_n(a_n^t|\bm{s}_n^t)
|a_n^t\in\mathcal{A}_n\}$ is the probability of CUAV $n$ choosing action $a_n^t$ at state $\bm{s}_n^t$ during time slot $t$ ($\pi_n(a_n^t|\bm{s}_n^t)\in[0,1]$). For this MARL process, each CUAV aims to find a strategy $\pi_n$ to maximize its average cumulative discounted reward, given the (implicit) impact of the adversary strategies of the other CUAVs.

It is known that without considering the influences of the other CUAVs' actions, the solution of (\ref{eq:8}) is a fixed point of the following Bellman equation, and an iterative search method can be used to find its solution,
\begin{equation}
\begin{aligned}
v_n(\bm{s}^0,\pi_n^*)=&\max_{a_n^t\in{\mathcal{A}_n}}
\{r_n^{t+1}(\bm{s}^t,a_n^t)\\
&+\gamma\sum_{\bm{s}^{t+1}}
P(\bm{s}^{t+1}|\bm{s}^t,a_n^t)v_n(\bm{s}^{t+1},\pi_n^*)\},
\end{aligned}
\label{eq:9}
\end{equation}
where $r_n^{t+1}(\bm{s}^t,a_n^t)$ is the instant reward of CUAV $n$ if it takes action $a_n^t$ over system state $\bm{s}^t$ at time slot $t$. $P(\bm{s}^{t+1}|\bm{s}^t,a_n^t)$ is the state transition probability as described in Section~\ref{sec31}.

Based on (\ref{eq:9}), the classical Q-learning method \cite{sutton2018reinforcement} can be adopted by each CUAV to approximate the solution to (\ref{eq:9}) by treating the adversary CUAVs as part of the stationary environment. Then, the Q-function is updated as
\begin{equation}
\begin{aligned}
q_n^{t + 1}\left( {{\bm{s}^t},{a_n^t}} \right) &\leftarrow \left( {1 - {\alpha ^t}} \right)q_n^t\left( {{\bm{s}^t},{a_n^t}} \right)\\
&+ {\alpha ^t}\left( {r_n^{t + 1}\left(\bm{s}^t,a_n^t \right) + \gamma \mathop {\max }\limits_a q_n^t\left( {{\bm{s}^{t+1}},a} \right)} \right),
\end{aligned}
\label{eq:10}
\end{equation}
where $q_n^{t+1}(\bm{s}^{t+1},a_n^t)$ is estimated state--action value at $t+1$ if CUAV $n$ takes action $a_n^t$ at state $\bm{s}^t$, ${\alpha ^t\in\left[0,1\right)}$ is the time-varying learning rate. It is proved in \cite{watkins1992q} that if $\sum_{t=1}^{\infty}\alpha^t=\infty$,  $\sum_{t=1}^{\infty}(\alpha^t)^2<\infty$ and the assumption of stationary environment holds, the iterative sequence based on Equation (\ref{eq:10}) converges to $q_n^{t+1}(\bm{s}^t,a_n^t)$ as each state is visited enough times.

Based on (\ref{eq:9}), we now consider the impact of the adversary policies on the performance of CUAV $n$ explicitly. Let $\bm{\pi}=(\pi_n,\bm{\pi}_{-n})$ and $\bm{a}_{-n}^t$ denote the actions of all the CUAVs except CUAV $n$ in time slot $t$. Then, (\ref{eq:8}) can be rewritten as follows,
\begin{equation}
\begin{aligned}
\max_{\pi_n}&v_n(\bm{s}^0,(\pi_n,\bm{\pi}_{-n}))=\\
&\max_{\pi_n}\sum_{t=0}^{+\infty}
\gamma^t\mathbb{E}(r_n^{t+1}(\bm{s}^t,(\pi_n,\bm{\pi}_{-n}))|{\bm{s}^0,(\pi_n,\bm{\pi}_{-n})}).
\end{aligned}
\label{eq:11}
\end{equation}

With (\ref{eq:11}), for $\forall\bm{s}^0\in\mathcal{S}$, each CUAV searches for the optimal $\pi_n$ to maximize its value function $v_n(\bm{s}^0,(\pi_n,\bm{\pi}_{-n}))$, given the stationary adversary policy $\bm{\pi}_{-n}$. The joint solution to~(\ref{eq:11}) for all $n\in\mathcal{N}$ leads to a Nash equilibrium (NE) solution, which can be mathematically defined as follows.

\begin{myDef}[Nash equilibrium \cite{zhang2021multi}]
An NE of the MG (as given in Definition~\ref{Definition_1})\\ $\langle\mathcal{N},\mathcal{S},\{\mathcal{A}_n\}_{n\in{\mathcal{N}}},
\mathcal{P},\{\mathcal{R}_n\}_{n\in{\mathcal{N}}},\gamma\rangle$ is a joint policy $\bm{\pi}^*=(\pi_n^*,\bm{\pi}_{-n}^*)$, s.t. for any $\bm{s}^0\in\mathcal{S}$ and $n\in\mathcal{N}$,
\begin{equation}
v_n(\bm{s}^0,(\pi_n^*,\bm{\pi}_{-n}^*))\geqslant v_n(\bm{s}^0,(\pi_n,\bm{\pi}_{-n}^*)),\forall \pi_n.
\label{eq:12}
\end{equation}
\label{Definition_2}
\end{myDef}

Although there always exists an NE for discounted MGs \cite{filar2012competitive}, guaranteeing the convergence to an NE through decentralized learning without exchanging the reward/policy information still remains an open problem. To tackle our considered problem in a decentralized manner, we leverage the idea of IL \cite{kaur2020energy}, and propose a Q-learning-based algorithm and a DDQN-based algorithm in Section \ref{sec4}. Fortunately, we are able to show the convergence of the proposed algorithms through numerical simulations in Section \ref{sec5}.

\section{Algorithm Design Based on Independent Learner}\label{sec4}
In this section, we introduce exploration strategy based on UCB-H, with which we develop two MARL algorithms in the framework of IL. The information exchanging overhead and execution complexity of the proposed algorithms are also discussed.

\subsection{UCB-H Strategy}
The main aim of introducing UCB-based action exploration strategy is to avoid the drawbacks of the traditional $\epsilon$-greedy strategy, which imposes no preference for the actions that are nearly greedy or particularly uncertain \cite{sutton2018reinforcement}.
The original UCB strategy is proposed for the multi-armed bandit scenario without discerning the underlying state evolution \cite{sutton2018reinforcement}:
\begin{equation}
a_n^t=\arg\max_a\left[{Q_n^t(a)+c\sqrt{\frac{\ln t}{N_n^t(a)}}}\right],
\label{eq:13}
\end{equation}
where $N_n^t(a)$ is the times that action $a$ has been selected prior to time slot $t$, and $c>0$ controls the degree of exploration. With~(\ref{eq:13}), actions with lower estimated values or that have already been selected frequently will be selected with decreasing frequency over time~\cite{sutton2018reinforcement}. For our concerned problem of channel selection, modification is needed to replace $N_n^t(a)$ by the times of selecting the state--action pair $(\bm{s}^t,\bm{a}^t)$.

For our studied problem, we introduce the UCB-H strategy to achieve a trade-off between action exploration and exploitation (cf. \cite{zhang2019multi,jin2018q}). Specifically, it also helps to balance a CUAV's strategy between preferring cooperation during sensing and incurring competition with more interference in channel access. Based on (\ref{eq:13}), the corresponding
\emph{Q}-value updating method now becomes (\ref{eq:14}) from (\ref{eq:10}):
\begin{equation}
\begin{aligned}
Q_n^{t + 1}\left( {{\bm{s}^t},{\bm{a}^t}} \right) &\leftarrow \left( {1 - {\alpha ^t}} \right)Q_n^t\left( {{\bm{s}^t},{\bm{a}^t}} \right) \\
&+ {\alpha ^t}\left( {r_n^{t + 1} + \mathop {\max }\limits_{a_n^{t + 1}} Q_n^t\left( {{\bm{s}^{t+1}},{\bm{a}^{t + 1}}} \right) + b^t} \right),
\end{aligned}
\label{eq:14}
\end{equation}
where
\begin{equation}
b^t=c\sqrt{\frac{H^3\ln(|\mathcal{S}||\mathcal{A}|T/p)}{N_n^t(\bm{s}^t,\bm{a}^t)}}.
\label{eq:15}
\end{equation}

In (\ref{eq:14}) and (\ref{eq:15}), $\alpha^t$ is learning rate that varies with time. $b^t$ is the confidence bonus indicating how certain the algorithm is about the current state--action pair. $N_n^t(\bm{s}^t,\bm{a}^t)$ is the times that state--action pair $(\bm{s}^t,\bm{a}^t)$ has been visited prior to time slot $t$. $T$ is the total number of time slots of the CUAV network operation. $p$ is an arbitrary small value to ensure that the total regret of the learning process is upper-bounded by $O(H^4|\mathcal{S}||\mathcal{A}|T\ln(|\mathcal{S}||\mathcal{A}|T/p))$ with probability $1-p$. $H$ is the steps in each episode of episodic Markov decision process (MDP) where $H=1$ in general MDP, i.e., our considered scenario \cite{zhang2019multi, jin2018q}.

\subsection{IL-Q-UCB-H Algorithm}
\begin{figure}[H]
\includegraphics[width=1.0\linewidth]{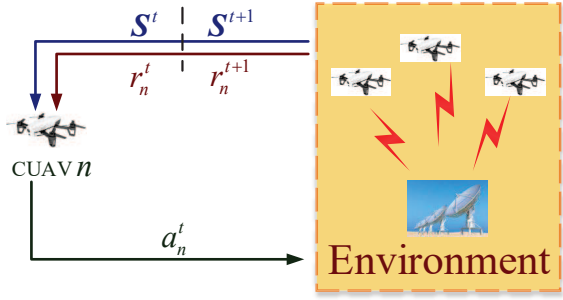}
\caption{$~$IL-Q-UCB-H of CUAV $n$ for joint sensing and access.\label{Figure4.}}
\end{figure}

By treating the other CUAVs as  part of the environment, the IL-Q-UCB-H algorithm can be developed based on standard Q-learning with UCB-H. This essentially approximates the original MARL problem in the MG by a group of single-agent RL problems, as shown in Fig.~\ref{Figure4.}.

For ease of generalization, we provide in (\ref{eq:16}) the traditional IL-Q algorithm that adopts $\epsilon$-greedy strategies for action selection. UCB-H can be conveniently incorporated into (\ref{eq:16}) by modifying the temporal difference term therein, as in (\ref{eq:14}).
\begin{equation}
\begin{aligned}
Q_n^{t + 1}\left( {{\bm{s}^t},{\bm{a}^t}} \right) &\leftarrow \left( {1 - {\alpha ^t}} \right)Q_n^t\left( {{\bm{s}^t},{\bm{a}^t}} \right) \\
&+ {\alpha ^t}\left( {r_n^{t + 1} + \gamma \mathop {\max }\limits_{a_n^{t + 1}} Q_n^t\left( {{\bm{s}^{t+1}},{\bm{a}^{t + 1}}} \right)} \right).
\end{aligned}
\label{eq:16}
\end{equation}

For $n\in\mathcal{N}$, we set the learning rate of IL-Q uniformly as \cite{cui2019multi}
\begin{equation}
\alpha^t=\frac{1}{(t+c_\alpha)^{\varphi_\alpha}},
\label{eq:17}
\end{equation}
where $c_\alpha>0$, $\varphi_\alpha\in\left(0.5,1\right]$. For either (\ref{eq:14}), the action update is obtained through tabular search:
\begin{equation}
a_n^{t+1}=\arg\max_{a_n^{t+1}}Q_n^{t+1}(\bm{s}^t,\bm{a}^{t+1}).
\label{eq:18}
\end{equation}

In summary, the IL-Q-UCB-H algorithm based on standard IL-Q learning is described in  \textbf{Algorithm~\ref{alg:1}}.

\begin{algorithm}[H]
\caption{IL-Q-UCB-H algorithm.}
\label{alg:1}
    \begin{algorithmic}[1]
    \STATE $\bm{Initialize}$: Set $t=0$, choose $p\in(0,1)$, $c>0$ $c_\alpha>0$, $\varphi_\alpha\in\left(0.5,1\right]$, and set the maximum time slots $T$;
    \FORALL{agent $n\in\mathcal{N}$}
        \STATE initialize $Q_n^t(\bm{s}^t,\bm{a}^t)=0$ and $\bm{s}^0$;
    \ENDFOR
    \WHILE{$t<T$}
        \FORALL{agent $n\in\mathcal{N}$}
            \STATE Update the learning rate $\alpha^t$ according to (\ref{eq:17});
            \STATE Select an action $a_n^t$ at $\bm{s}^t$ according to (\ref{eq:18});
            \STATE Take action $a_n^t$ to select channel for spectrum sensing and produce sensing decision $d_{n,m}^{t+1}$;
            \STATE Feedback sensing information $D_n^t=\{n,a_n^t,d_{n,m}^{t+1}\}$ on CCC;
            \STATE Receive sensing fusion decision $d_m^t$ according to (\ref{eq:7});
            \STATE Access channel based on sensing fusion decision, and receive reward $r_n^{t+1}$ according to (\ref{eq:2}) and observe $\bm{s}^{t+1}$;
            \STATE Update $Q_n^{t+1}(\bm{s}^t,\bm{a}^t)$ according to (\ref{eq:14});
        \ENDFOR
        \STATE $t=t+1$ and $\bm{s}^t\leftarrow \bm{s}^{t+1}$;
    \ENDWHILE
    \end{algorithmic}
\end{algorithm}

\subsection{IL-DDQN-UCB-H Algorithm}
\begin{figure}[H]
\includegraphics[width=1.0\linewidth]{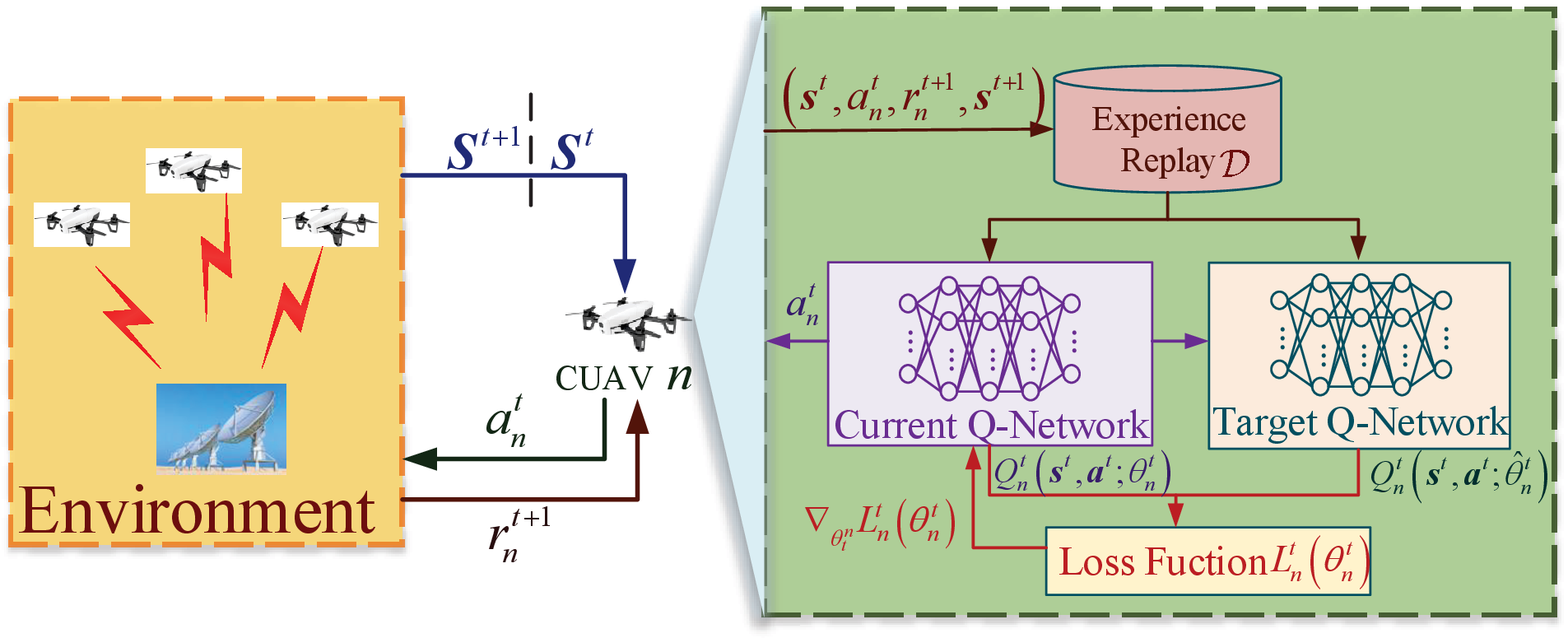}
\caption{$~$IL-DQN-UCB-H of CUAV $n$ for joint sensing and access.\label{Figure5.}}
\end{figure}

The proposed IL-Q-UCB-H algorithm requires each CUAV to construct a Q-table of size $|\mathcal{S}|\times|\mathcal{A}^n|$. Then, with the increasing number of PU channels, the IL-Q-UCB-H algorithm faces the curse of dimensionality. To handle such a problem, we adopt the framework of DDQN \cite{van2016deep} for value space approximation with deep neural networks which replace the IL-Q-UCB-H algorithm with the IL-DDQN-UCB-H algorithm. Compared with the vanilla DQN algorithm, the core of the IL-DDQN-UCB-H algorithm decomposes the maximization operation into a neural network for action selection and a target neural network for action evaluation \cite{van2016deep}. The main functional components \cite{li2020deep} are illustrated in Fig.~\ref{Figure5.}, and each component is described in detail as follows.

\emph{Input Layer}: The input of DDQN is a vector of size $(2M+1)$, corresponding to the system state $\bm{s}^t=(s_0^t,\ldots,s_M^t,o_1^t,\ldots,o_M^t)$ in time slot $t$, where the first $M+1$ value corresponds to the number of CUAVs that select each PU channel to sense, or does not select any PU channel, and the last $M$ value indicates the occupancy state of each PU channel,~respectively.

\emph{Output Layer}: The output of DDQN is a vector of size $(M+1)$, corresponding to the \emph{Q}-value estimation of all optional actions given the current system state, i.e.,
$Q_n^t=[Q_{n,0}^t,Q_{n,1}^t,\ldots,Q_{n,M}^t]$.

\emph{Experience Replay}: In DDQN, the experience replay component stores the accumulated samples in history in the form of experience tuples $(\bm{s}^t,a_n^t,r_n^{t+1},\bm{s}^{t+1})$ which are composed of the current state $\bm{s}^t$, action $a_n^t$, reward $r_n^{t+1}$, and the next state $\bm{s}^{t+1}$. During the learning process, the agent randomly samples a batch of experience tuples of length $B$ from the experience replay to fit the deep network to the \emph{Q}-values, aiming to eliminate the temporal correlation of historical samples.

\emph{Current Q-Network}: The current Q-network (i.e., Q-table fitting deep neural network) realizes the mapping of the input state $\bm{s}^t$ to the corresponding \emph{Q}-value $Q_n^{t+1}(\bm{s}^t,\bm{a}^{t+1};\theta_n^t)$ of each action $a_n^t$, where $\theta_n^t$ is the parameters of the current Q-network. The experience tuples are mainly used to train the current Q-network to update its own set of parameters $\theta_n^t$ until convergence. After training, an action will be selected based on the output \emph{Q}-values.

\emph{Target Q-Network}: The target Q-network has the same structure as the current Q-network, also with the same initial parameters. The output target \emph{Q}-value $Q_n^{t+1}(\bm{s}^t,\bm{a}^{t+1};\hat{\theta}_n^t)$ is mainly used to supervise the iterative training of the current Q-network, where $\hat{\theta}_n^t$ is the parameters of the target Q-network. In DDQN, $\hat{\theta}_n^t$ is updated after a fixed rounds $F$ of training. It directly assigns the value of $\theta_n^t$  to $\hat{\theta}_n^t$, which is known as the fixed Q-targets in~DDQN.

\emph{Action selection strategy}: To prevent the actions falling into the local optimum during the period of unconverged deep neural network training stage, the greedy strategy is introduced during action selection (cf. (\ref{eq:18})),
\begin{equation}
a_n^{t+1}=\arg\max_{a_n^{t+1}}Q_n^{t+1}(\bm{s}^t,\bm{a}^{t+1};\theta_n^t).
\label{eq:19}
\end{equation}

\emph{Loss Function}: The loss function used in training the current Q-Network is defined as~follows:
\begin{equation}
L_n^t(\theta_n^t)=\frac{1}{B}\sum_{i=1}^B(y_{n,i}
-Q_{n,i}^t(\bm{s}^t,\bm{a}^t;\theta_n^t))^2,
\label{eq:20}
\end{equation}
where $B$ is the batch size and $y_{n,i}$ is the target \emph{Q}-value. With UCB-H, the updating method of the target \emph{Q}-value is
\begin{equation}
a_n^{t,max}=\arg\max_{a_n^t}Q_{n,i}^t(\bm{s}^t,\bm{a}^t;\theta_n^t),
\label{eq:21}
\end{equation}
with
\begin{equation}
y_{n,i}=r_{n,i}^{t+1}
+\gamma\max_{a_n^t}Q_{n,i}^t(\bm{s}^{t+1},a_n^{t,max};\hat{\theta}_n^t)+b^t.
\label{eq:22}
\end{equation}

We note that the loss function is a mean square error between the output \emph{Q}-value of the target Q-network and that of the current Q-network. After receiving the value of the loss function, the gradient descent method is used to update $\theta_n^t$ iteratively, i.e.,
\begin{equation}
\theta_n^{t+1}\leftarrow\theta_n^t+
\zeta\nabla_{\theta_n^t}L_n^t(\theta_n^t)
\label{eq:23}
\end{equation}
with a learning rate $\zeta$. The gradient $\nabla_{\theta_n^t}L_n^t(\theta_n^t)$ is calculated following (\ref{eq:24})
\begin{equation}
\nabla_{\theta_n^t}L_n^t(\theta_n^t)=\nabla_{\theta_n^t}
\left[\frac{1}{B}\sum_{i=1}^B\left(y_{n,i}
-Q_{n,i}^t(\bm{s}^t,\bm{a}^t;\theta_n^t)\right)^2\right].
\label{eq:24}
\end{equation}

For the considered CUAV network, the framework of the IL-DDQN-UCB-H algorithm is given in \textbf{Algorithm~\ref{alg:2}} based on the aforementioned functional components.

\begin{algorithm}
\caption{IL-DDQN-UCB-H Algorithm.}
\label{alg:2}
    \begin{algorithmic}[1]
    \STATE $\bm{Initialize}$: Set $t=0$, choose $\gamma\in\left[0,1\right)$, $p\in(0,1),c>0$, and set the maximum time slots $T$, experience replay size $C$, batch size $B$, target Q-Network update period $F$, DDQN learning rate $\zeta$;
    \FORALL{agent $n\in\mathcal{N}$}
        \STATE Randomly initialize the current Q-network parameters $\theta_n^t$, target Q-network parameters $\hat{\theta}_n^t$ and $\bm{s}^0$;
    \ENDFOR
    \WHILE{$t<T$}
        \FORALL{agent $n\in\mathcal{N}$}
            \STATE Select an action $a_n^t$ at $\bm{s}^t$ according to (\ref{eq:19});
            \STATE Take action $a_n^t$ to select channel for spectrum sensing and produce sensing decision $d_{n,m}^{t+1}$;
            \STATE Feedback sensing information $D_n^t=\{n,a_n^t,d_{n,m}^{t+1}\}$ on CCC;
            \STATE Receive sensing fusion decision $d_m^t$ according to (\ref{eq:7});
            \STATE Access channel based on sensing fusion decision, and receive reward $r_n^{t+1}$ according to (\ref{eq:2}) and observe $\bm{s}^{t+1}$;
            \STATE Store $(\bm{s}^t,a_n^t,r_n^{t+1},\bm{s}^{t+1})$ into experience replay;
            \IF{$t>C$}
                \STATE Remove the old experience tuples from experience replay;
            \ENDIF
            \STATE Randomly select a batch size $B$ experience tuples $(\bm{s}^t,a_n^t,r_n^{t+1},\bm{s}^{t+1})$ from experience replay;
            \STATE Calculate loss function  $L_n^t(\theta_n^t)$ according to (\ref{eq:20}) and (\ref{eq:24});
            \STATE Update parameter $\theta_n^t$ according to (\ref{eq:23});
            \IF{$t\bmod F=0$}
                \STATE $\hat{\theta}_n^t\leftarrow\theta_n^t$;
            \ENDIF
        \ENDFOR
        \STATE $t=t+1$ and state $\bm{s}^t\leftarrow\bm{s}^{t+1}$;
    \ENDWHILE
    \end{algorithmic}
\end{algorithm}

\subsection{Algorithm Complexity Analysis}
\begin{itemize}
    \item \emph{IL-Q-UCB-H algorithm}
 : Since each CUAV executes the IL-Q-UCB-H algorithm independently, its information interaction overhead is mainly caused by broadcasting its own sensing decision information. The amount of information interaction increases linearly with the increase of CUAVs. For algorithm execution, each CUAV needs to store a Q-table of size $N\cdot 2^M(M+1)^N$ according to the number of states and actions. It increases exponentially with the numbers of CUAVs and PU channels. The computational cost for each CUAV is dominated by the linear update of the Q-table and the search for the optimal action, which are both of constant time complexity.
    \item \emph{IL-DDQN-UCB-H algorithm}: The cost of information exchange is the same as the IL-Q-UCB-H algorithm. For algorithm execution, since a deep neural network is used to fit the \emph{Q}-values, the storage cost mainly depends on the structure of the deep neural network. Since the IL-DDQN-UCB-H algorithm involves updating two Q-networks, the computational complexity is dependent of the neural network structure (i.e., the network parameters) at the training stage.
\end{itemize}

\section{Simulation and Analysis}\label{sec5}
In this section, the performance of the proposed algorithms is evaluated in the same CUAV network through numerical simulations. Specifically, the experiments are carried out with respect to several indicators, including the average reward, sensing accuracy, and channel utilization. 
The average reward is evaluated as the average instant reward of all the CUAVs, $\bar r^{t+1}=N^{-1}\sum^N_{n=1}r_n^{t+1}$. The sensing accuracy is evaluated as $acc=(N_{acc}^t/M)\times100\%$, where $N_{acc}^t$ is the number of PU channels over which the sensing fusion produces correct observation of the channel states. The channel utilization is evaluated as $uti=(N_{uti}^t/M)\times100\%$ where $N_{uti}^t$ is the number of PU  channels selected by CUAVs in time slot $t$. The main parameters used throughout the simulations are given in Table~\ref{tab:1}. The binary Markov model for PU activities are randomly initialized as $(\alpha_m,\beta_m),\forall m=1,\ldots,M$. The hyperparameters of all the RL algorithms are given in Table~\ref{tab:2}. The learning rate $\alpha^t$ is initialized as 0.9.

\begin{table}[htbp]
\centering
\captionsetup{font={footnotesize}}
\caption{\,Simulation parameters.}
\begin{tabular}{cc}
  \hline
  \toprule Parameters & Value\\
  \midrule
  PU channels             $M$        & 5 \\
  CUAV number             $N$        & 4, 5, 10\\
  Channel bandwidth       $B_m$      & 50$\sim$100~MHz \\
  False alarm probability $P_f$      & 0.1 \cite{zhang2019multi} \\
  Detection probability   $P_d$      & 0.9 \\
  Transmission power      $P_t$      & 23~dBm \cite{cui2019multi} \\
  Sensing time            $\tau_s$   & 0.1~ms\\
  Transmission time       $\tau_t$   & 0.5~ms \\
  Weights of sensing/access cost    $\eta,\mu$ & 0.01, 0.05 \\
  \bottomrule
  \hline
\end{tabular}
\label{tab:1}
\end{table}

\begin{table}[h]
\centering
\captionsetup{font={footnotesize}}
\caption{\,Hyperparameters of the RL algorithms.}
\begin{tabular}{cc}
  \hline
  \toprule Hyper-parameters & Value\\
  \midrule
  Greedy rate             $\epsilon$   & 0.1 \\
  Discount factor         $\gamma$     & 0.9 \\
  Parameters of the learning rate $c_\alpha,\varphi_\alpha$& 0.5, 0.8 \cite{cui2019multi} \\
  Parameters of UCB-H     $p,c$        & 0.01, 2 \cite{zhang2019multi} \\
  Parameters of CNN                    & (2, 2, 10)\\
  Activation function                  & ReLu \cite{cai2020coordination}\\
  Optimizer                            & Adam \cite{kingma2014adam}\\
  Batch size $B$                       & 64 \\
  Target Q-Network update period $F$   & 100 \\
  Experience replay size  $C$          & 20,000 \\
  \bottomrule
  \hline
\end{tabular}
\label{tab:2}
\end{table}

\begin{figure}
\subfigure[Evolution of the average reward.]{\includegraphics[width=1.0\linewidth]{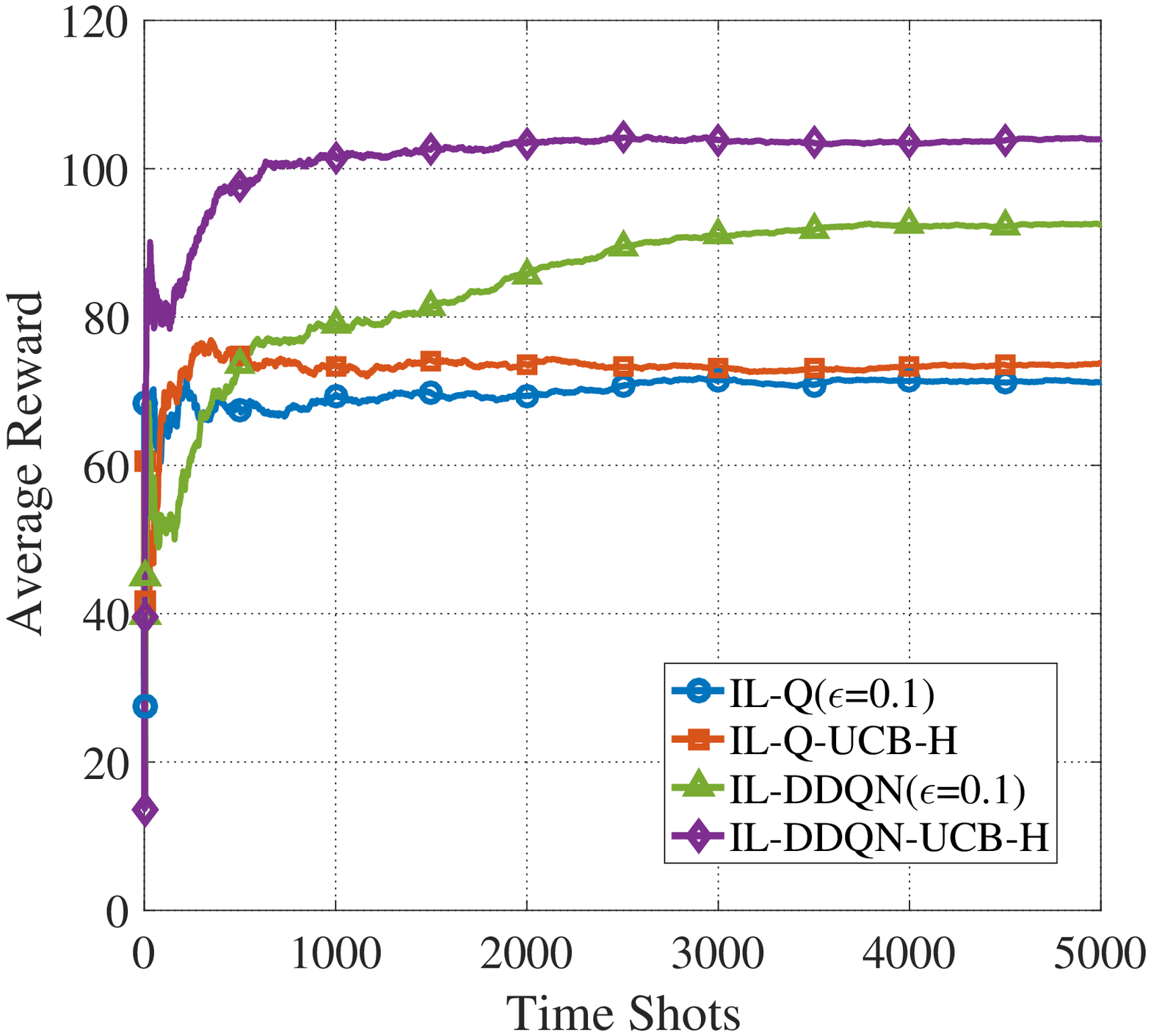}}
\subfigure[Evolution of the sensing accuracy.]{\includegraphics[width=1.0\linewidth]{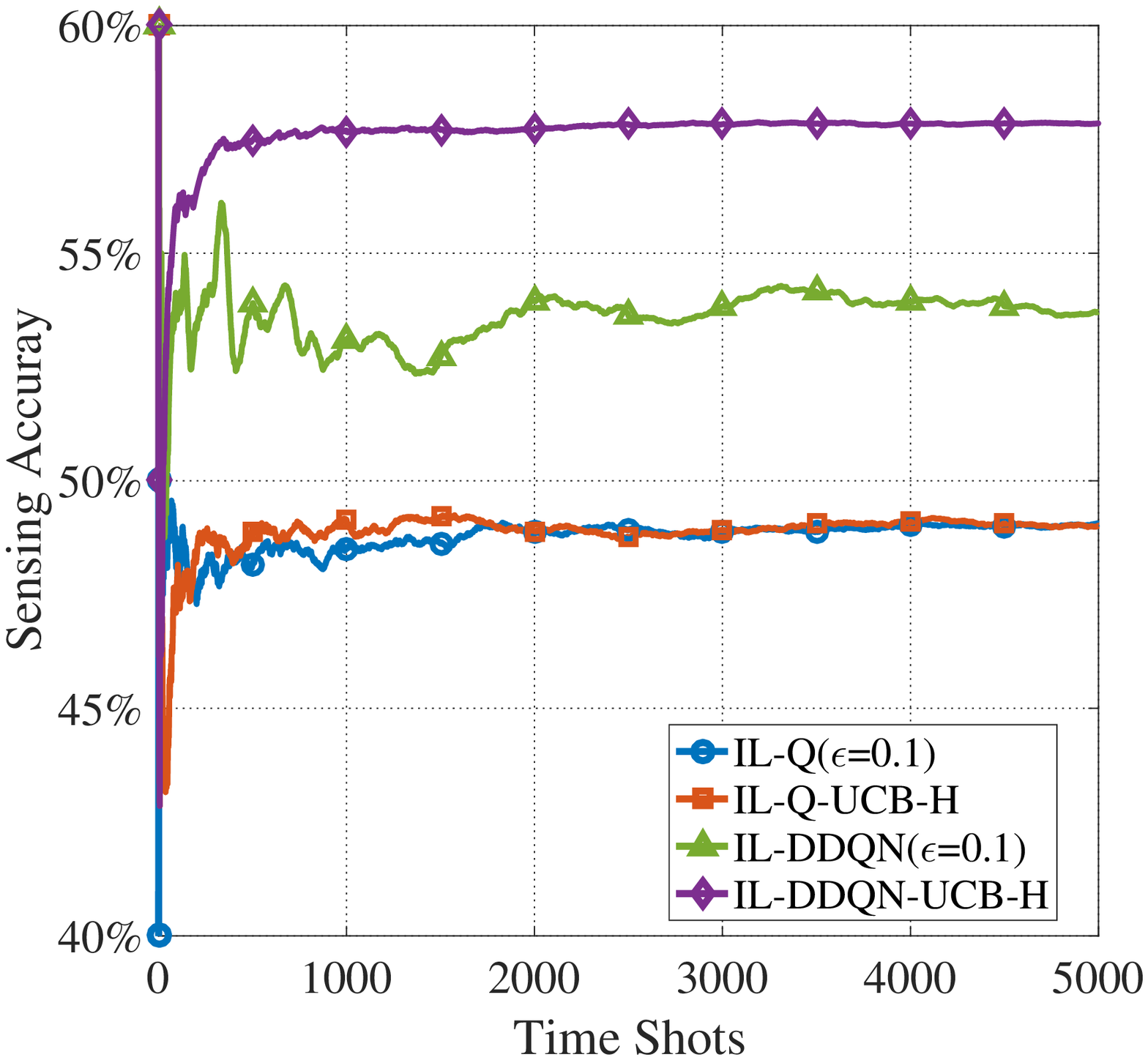}}
\caption{$~$Evolution of the average reward and the sensing accuracy with training ($N=4,M=5$).\label{Figure6.}}
\end{figure}

\begin{figure}
\subfigure[Evolution of the average reward.]{\includegraphics[width=1.0\linewidth]{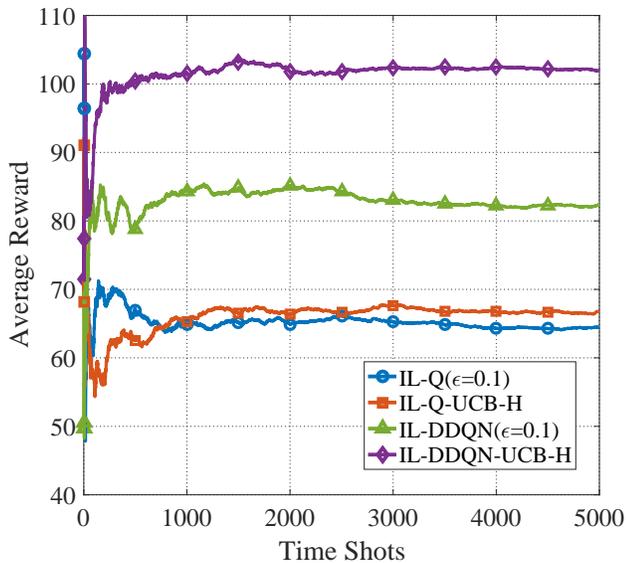}}
\subfigure[Evolution of the sensing accuracy.]{\includegraphics[width=1.0\linewidth]{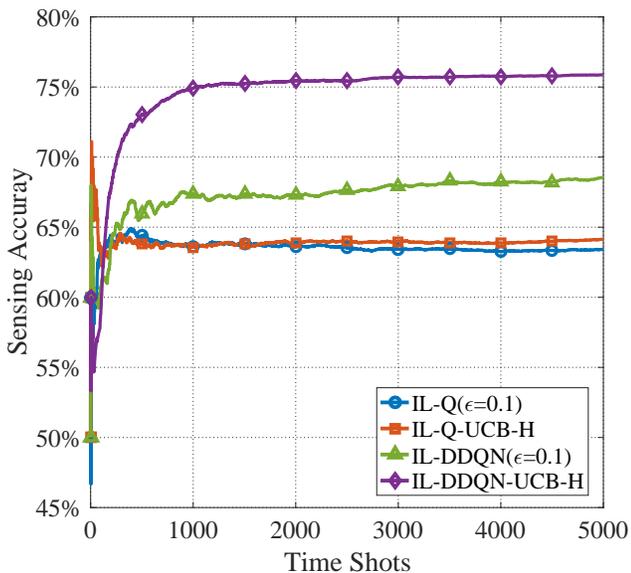}}
\caption{$~$Evolution of the average reward and the sensing accuracy with training ($N=6,M=5$).\label{Figure7.}}

\end{figure}

To demonstrate that the proposed algorithms are able to handle the network congestion, the simulations in Fig.~\ref{Figure6.} and Fig.~\ref{Figure7.} evaluate the average reward and sensing accuracy for two cases of $N=4,M=5$ and $N=6, M=5$.

\unskip

\unskip
We observe from Fig.~\ref{Figure6.}(a) that all of the four algorithms are able to converge with sufficient training epochs. We note that the two IL-DDQN algorithms are able to obtain higher average reward than the two IL-Q algorithms. The reason lies in that DDQN not only reduces the correlation of sampled data, but also prevents overfitting to handle the excessive state--action space more efficiently. At the same time, the UCB-H-enabled algorithms are able to achieve higher average rewards than their $\epsilon$-greedy counterparts. This indicates that the UCB-H strategy is able avoid the performance degradation caused by the randomness due to $\epsilon$-greedy exploration and the local optimality caused by insufficient exploration when using myopic strategy to select actions.

Fig. \ref{Figure6.}(b) evaluates the sensing accuracy of the four algorithms with $N=4,M=5$. It can be seen that, similar to Fig. \ref{Figure6.}(a), the performance of IL-Q-UCB-H and IL-DDQN-UCB-H is also better than $\epsilon$-greedy IL-Q and IL-DDQN. In addition, the $\epsilon$-greedy-enabled algorithms fluctuate more severely in the early stage of training. The reason is that the \emph{Q}-values using the $\epsilon$-greedy strategy bear little difference at the early stage, and this makes the agents select actions randomly. The UCB-H-enabled algorithms are relatively smooth in the early stage of training, thanks to the confidence bonus, which makes the \emph{Q}-values discernible. In summary, Fig.~\ref{Figure6.} shows that the proposed IL-DDQN-UCB-H algorithm is able to achieve the best performance, in terms of the average reward and the sensing accuracy, when the number of CUAVs are less than that of PUs and the congestion does not~exist.

Fig.~\ref{Figure7.} shows the performance in terms of the average reward and sensing accuracy of the four algorithms with $N=6,M=5$. As can be seen from the figure, the UCB-H-enabled algorithms are able to achieve better performance in the condition of congestion. In addition, comparing Fig.~\ref{Figure6.}(b) and Fig.~\ref{Figure7.}(b), we note that when there are more CUAVs, the sensing accuracy rate can be increased by 10\% to 15\%. This demonstrates the efficiency of the cooperative sensing mechanism.

\begin{figure}
\subfigure[Evolution of the average reward.]{\includegraphics[width=1.0\linewidth]{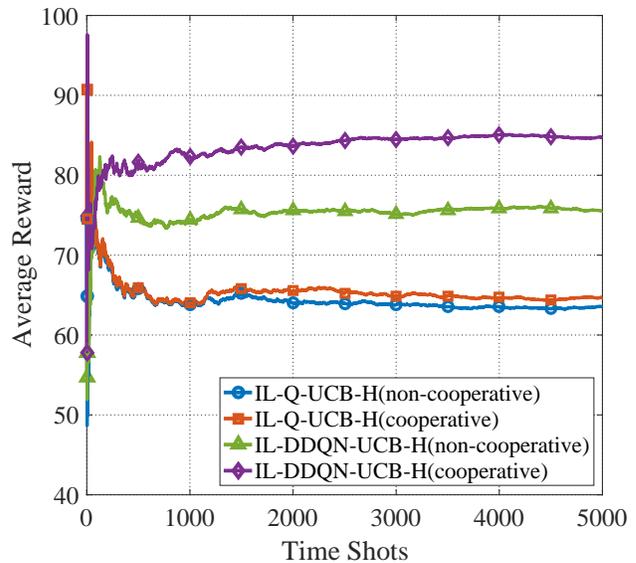}}
\subfigure[Evolution of the sensing accuracy.]{\includegraphics[width=1.0\linewidth]{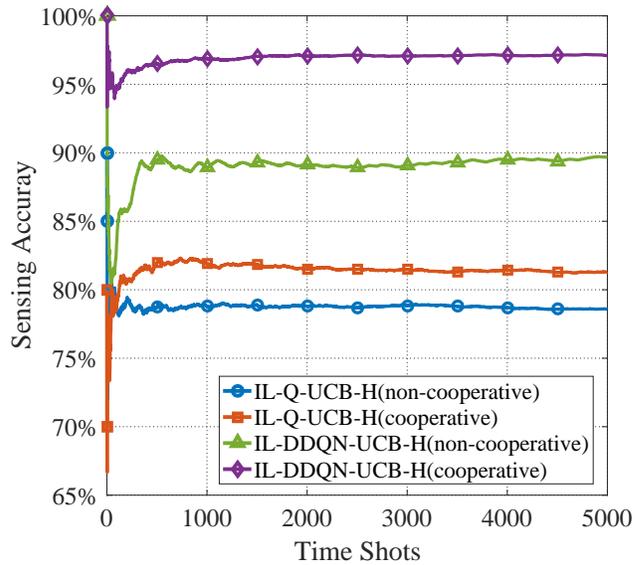}}
\caption{$~$Evolution of the average reward and the sensing accuracy of the proposed algorithms in cooperative and non-cooperative scenarios ($N=10,M=5$).\label{Figure8.}}
\end{figure}

A further illustration of the trade-off between the sensing accuracy and network congestion is provided by Fig.~\ref{Figure8.} with $N=10$. It can be seen that the performance of CUAV cooperation is significantly better than that of non-cooperation. In particular, the sensing accuracy of the IL-DDQN-UCB-H algorithm in the cooperative scenario can reach 97\%. At the same time, the achieved average reward of cooperation is less than the cases of $N=4$ or $N=6$, which indicates that the improved accuracy may not fully compensate the degradation of transmission due to congestion.

\begin{figure}[]
\includegraphics[width=1.0\linewidth]{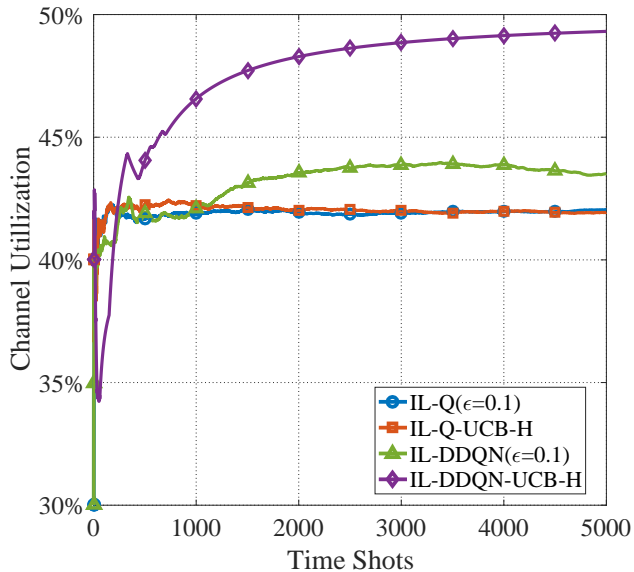}
\caption{$~$Evolution of the channel utilization of four algorithms ($N=10,M=5$).\label{Figure12.}}
\end{figure}

\unskip
Considering the cases where some CUAVs do not select a channel for sensing and access in every time slot, another simulation is performed with channel utilization as an performance indicator. Fig. \ref{Figure12.} shows the channel utilization performance of the four algorithms. It can be seen that the four algorithms can achieve a channel utilization of more than 42\%, especially the IL-DDQN-UCB-H algorithm which has a channel utilization of 49\%. It shows that the proposed cooperative sensing and access algorithms can find idle PU channels in time and significantly improve the channel utilization.

\begin{figure}[H]
\includegraphics[width=1.0\linewidth]{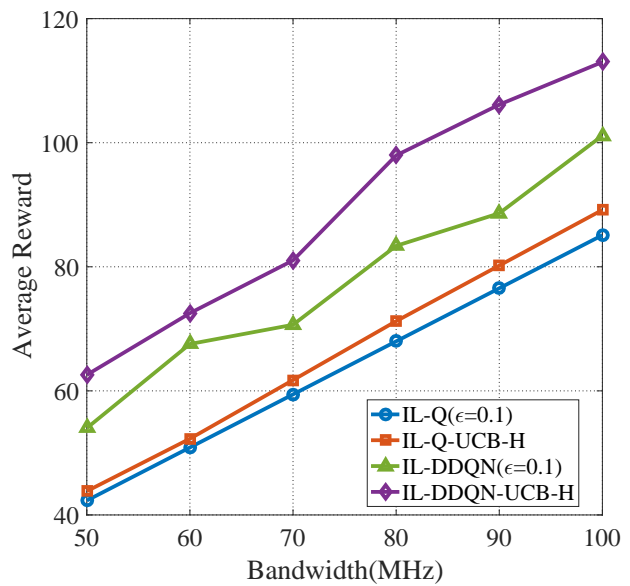}
\caption{$~$Evolution of the average reward of four algorithms with different bandwidths ($N=4,M=5$).\label{Figure13.}}
\end{figure}

We note from Section \ref{sec32} that there are four situations for CUAVs to sense and access PU channels. The obtained reward is dependent on the channel bandwidth in these four situations. This is mainly reflected in the spectrum sensing cost and the available data transmission volume (utility). By the definition of the reward function, the spectrum sensing cost $-E_{s,n}^{t+1}$ is a negative reward and has a negative correlation with the channel bandwidth, while $R_n^{t+1}>0$ with a positive correlation with the channel bandwidth. As the channel bandwidth increases, the absolute values corresponding to the cost and utility will also increase, resulting in a decrease in the system reward. The simulation analyzes the relationship between the average reward and PU channel bandwidth. PU channel bandwidth is taken as $B_m\in\{50,60,70,80,90,100\}$~MHz and the result is shown in Fig.~\ref{Figure13.}. It can be found that as the channel bandwidth increases, the system average reward also increases. This indicates that the cost due to sensing a larger bandwidth can be compensated by the utility gained from channel utilization. Namely, choosing a PU channel with a large channel bandwidth to construct a set of candidate sensing channels generally leads to better performance of the CUAV network.

\begin{figure}[H]
\includegraphics[width=1.0\linewidth]{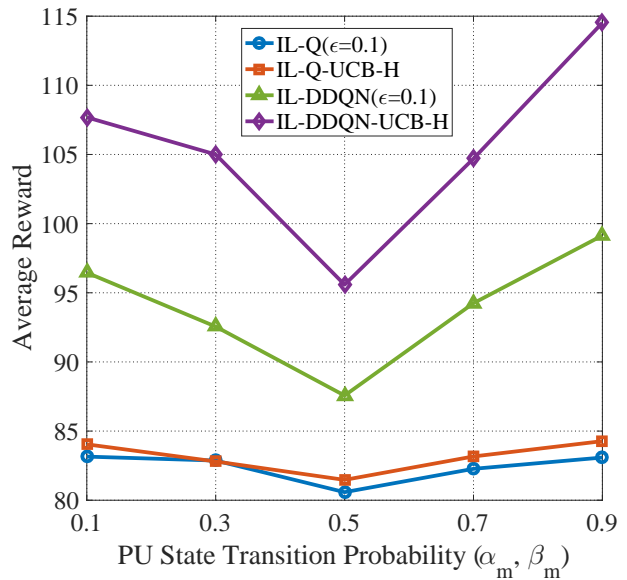}
\caption{{$~$Evolution of the average reward of four algorithms with different PU state transition probabilities ($N=4,M=5$)}.\label{Figure14.}}
\end{figure}

The average reward of the four algorithms under different PU channel state transition probabilities is analyzed with $(\alpha_m,\beta_m)$ varying as $\alpha_m=\beta_m\in\{0.1,0.3,0.5,0.7,0.9\}$. \mbox{Fig.~\ref{Figure14.}} shows that when the state transition probabilities $(\alpha_m,\beta_m)$ increase from 0.1 to 0.5, the average reward decreases. Comparatively, when it gradually increases from 0.5 to 0.9, the average reward increases. As shown in Fig. \ref{Figure2.}, the randomness of PU channel state is small when $(\alpha_m,\beta_m)$ is either very large or small. In this situation, the CUAVs estimate PU channel states more accurately based on the  historical experience, and greater rewards can be obtained based on this decision. However, PU channel state transition is highly random when $(\alpha_m,\beta_m)$ is about 0.5. In this situation, the reward will decrease based on the historical experience of the CUAVs and so will the sensing accuracy.

\section{Conclusion}\label{sec6}
In this paper, the problem of joint spectrum sensing and channel access for a CUAV communication network in a time-varying radio environment was studied. In a situation where the information about the primary network dynamics is not known in advance, a competition--cooperation protocol framework was proposed for CUAVs to implicitly cooperate over the channels to sense and access. An MG-based model was introduced to translate the centralized one-shot network optimization problem into a group of MARL problems that locally optimize the cumulative sensing--transmission reward of each CUAV. To avoid excessive information exchange overhead for channel cooperation, an independent Q-learning algorithm and an independent DDQN algorithm were proposed to approximate the equilibrium strategies of the MG. The proposed learning algorithms were improved with the UCB-H-based action--exploration strategy. Numerical simulation results showed that the proposed algorithms can increase the system average reward, sensing accuracy, and channel utilization efficiently.





\ifCLASSOPTIONcaptionsoff
  \newpage
\fi

\bibliographystyle{IEEEtran}
\bibliography{Myref}





\end{document}